\documentclass{aastex}
\usepackage{amsbsy}
\usepackage{amsfonts}
\usepackage{amssymb}
\usepackage{longtable}
\usepackage{lscape}

\usepackage{graphicx}
\usepackage{graphics}
\usepackage{dcolumn}
\usepackage{bm}
\begin{document}


\title{Measurements of Anisotropic Ion Temperatures, Non-Thermal Velocities, and Doppler Shifts in a Coronal Hole}
\author{M. Hahn\altaffilmark{1} and D. W. Savin\altaffilmark{1}}

\altaffiltext{1}{Columbia Astrophysics Laboratory, Columbia University, MC 5247, 550 West 120th Street, New York, NY 10027 USA}

\date{\today}
\begin{abstract}

	We present a new diagnostic allowing one to measure the anisotropy of ion temperatures and non-thermal velocities as well as Doppler shifts with respect to the ambient magnetic field. This method provides new results, as well as independent test for previous measurements obtained with other techniques. Our spectral data come from observations of a low latitude, on-disk coronal hole. A potential field source surface model was used to calculate the angle between the magnetic field lines and the line of sight for each spatial bin of the observation. A fit was performed to determine the line widths and Doppler shifts parallel and perpendicular to the magnetic field. For each line width component we derived ion temperatures $T_{\mathrm{i},\perp}$ and $T_{\mathrm{i},\parallel}$ and non-thermal velocities $v_{\mathrm{nt},\perp}$ and $v_{\mathrm{nt},\parallel}$. $T_{\mathrm{i},\perp}$ was cooler than off-limb polar coronal hole measurements. $T_{\mathrm{i},\parallel}$ is consistent with a uniform temperature of $1.8 \pm 0.2 \times 10^{6}$~K for each ion. Since parallel ion heating is expected to be weak, this ion temperature should reflect the proton temperature. A comparison between our results and others implies a large proton temperature gradient around 1.02~$R_{\sun}$. The non-thermal velocities are thought to be proportional to the amplitudes of various waves. Our results for $v_{\mathrm{nt},\perp}$ agree with Alfv\'en wave amplitudes inferred from off-limb polar coronal hole line width measurements. Our $v_{\mathrm{nt},\parallel}$ results are consistent with slow magnetosonic wave amplitudes inferred from Fourier analysis of time varying intensity fluctuations. Doppler shift measurements yield outflows of $\approx 5$~$\mathrm{km\,s^{-1}}$ for ions formed over a broad temperature range. This differs from other studies which found a strong Doppler shift dependence on formation temperature.  	
	
\end{abstract}

\maketitle
	
\section{Introduction} 

	Wave- and turbulence-driven models of coronal heating and solar wind acceleration propose that the necessary energy is carried to larger heights via plasma waves driven by agitation in and below the photosphere \citep{Cranmer:SSR:2002}. Attempts to theoretically describe the dissipation of these waves and how that energy is converted into particle heating represent a major area of research in solar physics. Observations of the temperature, wave properties, and flow velocities of the plasma can constrain these models. In the corona, where solar wind acceleration begins, emission line profiles can be used to infer many of the initial properties of the solar wind. Further out, in-situ measurements near 1~AU provide a detailed description of the plasma. 
	
	The bulk motions induced by waves are expected to be anisotropic with respect to the ambient magnetic field. The two main types of wave in the solar corona are Alfv\'en waves and slow magnetosonic waves. Alfv\'en waves are transverse waves and can be observed spectroscopically through non-thermal line broadening perpendicular to the magnetic field. There is strong evidence that Alfv\'en waves exist throughout the Sun from the chromosphere \citep{DePontieu:Sci:2007} into the corona \citep{Tomczyk:Sci:2007} and into the solar wind \citep{Belcher:JGR:1971}. Spectroscopic measurements indicate that these waves can be important for coronal heating \citep{Hahn:ApJ:2012, Bemporad:ApJ:2012}. 
	
	On the other hand, slow magnetosonic waves cause non-thermal broadening in the direction parallel to the magnetic field. These are essentially sound waves, but they are modified in the solar corona by the strong magnetic pressure which constrains the fluid motion of the wave to be nearly along the magnetic field. Because they are compressive, these waves may also be observed as time varying line intensity oscillations. Intensity fluctuations consistent with slow magnetosconic waves have been reported in a number of observations \citep[e.g.,][]{Banerjee:SSR:2011}. 
	
	It is also known that the ion temperature can be anisotropic in the solar wind. For example, in-situ measurements near 1~AU indicate that proton temperatures are anisotropic with $T_{\mathrm{p},\perp} > T_{\mathrm{p},\parallel}$ \citep{Marsch:JGR:2004}. Such anisotropic heating can be caused by cyclotron resonance with high frequency waves \citep{Cranmer:ApJ:1999} or by stochastic heating from large amplitude fluctuations transverse to the magnetic field \citep{Chandran:ApJ:2010}. 
		
	However, closer to the Sun the ability to detect ion temperature anisotropy and distinguish between Alfv\'en and magnetosonic waves through emission line profiles has been limited because spectroscopic observations usually only detect one component of the line width. Typical line width measurements in off-limb spectra of coronal holes or quiet Sun regions are primarily sensitive to perpendicular broadening. This is because the line of sight is nearly perpendicular to the magnetic field lines at the point where the line of sight passes closest to the Sun. Since the electron density $n_{\mathrm{e}}$ drops exponentially with height, and line intensity is proportional to $n_{\mathrm{e}}^2$ \citep[e.g.,][]{Hahn:ApJ:2010}, any varations of the angle along the line of sight has little effect on the measured width. Thus, such measurements provide a useful diagnostic for Alfv\'en waves and perpendicular ion temperatures, but provide no information about sonic waves or the parallel ion temperature. 
	
	Several attempts have been made to overcome these limitations. \citet{Kohl:ApJ:1998} observed a polar coronal hole above 1.5~$R_{\sun}$ with the Ultraviolet Coronagraph Spectrometer \citep[UVCS;][]{Kohl:SolPhys:1995}. By combining line width and Dopper dimming diagnostics for an O~\textsc{vi} line they were able to measure the perpendicular ion temperature $T_{\mathrm{i},\perp}$ as well as put upper and lower bounds on $T_{\mathrm{i},\parallel}$. They determined that $T_{\mathrm{i},\perp} > T_{\mathrm{i},\parallel}$ down to at least $2.2$~$R_{\sun}$. 
	
	An alternative method to separate parallel and perpendicular broadening has been to measure the variation of line widths from the center of the solar disk to the limb. The principle of this method is that, on average, the line of sight looks down the field lines of large scale magnetic loops at the center of the disk, while near the limb the observations tend to look across the field lines. The results from such studies have been mixed. \citet{Feldman:ApJ:1976} found no change in the line width, whereas center-to-limb broadening has been found by \citet{Roussel:MNRAS:1979} and \citet{Erdelyi:AA:1998}. Even if a clear trend could be discerned, the precise separation of the parallel and perpendicular broadening would remain ambiguous because the magnetic field direction is not known except in an average sense.

	Flow velocities provide another constraint on solar wind models that complement the temperature and non-thermal velocity data. The flow of material from the Sun into the solar wind can be inferred from Doppler shifts. The ability to make precise measurements of the Doppler shift depends on the precision with which the rest wavelength on the detector is known for the observed spectral line. Since most spectrometers do not carry an onboard calibration lamp it is usually not possible to use the laboratory wavelength for this. One way to determine the rest wavelength is to study the center-to-limb variation of the line center \citep{Peter:ApJ:1999}. This method assumes that at the limb any non-perpendicular motions cancel out when summed along the line of sight. The disadvantage of this method is that the Doppler shift is measured only in an average sense. Another method is to assume that there are no flows in the chromosphere and measure all wavelengths with respect to a chromospheric line. Since there probably are flows in the chromosphere this method is only accurate when the velocity in the corona is large relative to any chromospheric velocity. 
			
	Here, we describe a new diagnostic which allows one to measure both the parallel and perpendicular components of a line width and also the flow velocity. We use an observation with the Extreme ultraviolet Imaging Spectrometer \citep[EIS;][]{Culhane:SolPhys:2007} taken of a coronal hole at a relatively low latitude. The slit spanned a region where there was a large variation in the angle between the line of sight and the magnetic field. A Potential Field Source Surface (PFSS) model was used to trace the magnetic field lines passing through each spatial bin of the EIS data. As we explain later, this allowed us to calculate both the line width and Doppler shift as a function of the inclination angle between the line of sight and the magnetic field and thereby separate their parallel and perpendicular components. For the line widths, we analyzed the two components separately to estimate the partition between thermal and non-thermal broadening. From the Doppler shifts we determined the bulk flow by assuming that there is no motion perpendicular to the field lines. As we explain below, this allows us to fit for the rest wavelength and thereby infer the parallel velocity from the Doppler shift relative to that wavelength.
	
	The rest of this paper is organized as follows: Section~\ref{sec:ana} describes our analysis including details of the observation, fitting of spectral lines to determine the line widths, the PFSS magnetic field model, and separation of the parallel and perpendicular components. Results are presented in Section~\ref{sec:res}. In Section~\ref{subsec:dentemp} we characterize the observed region in terms of the electron temperature and density. Sections~\ref{subsec:perp} and \ref{subsec:par} then discuss the ion temperatures and non-thermal velocities extracted from the perpendicular and parallel line widths and compare these results to measurements using other methods. Our Doppler shift results are presented in Section~\ref{subsec:dopp}. Section~\ref{sec:con} summarizes our conclusions. 

\section{Analysis} \label{sec:ana}
\subsection{Observation}	\label{subsec:obs}

\begin{figure}
\centering \includegraphics[width=0.9\textwidth]{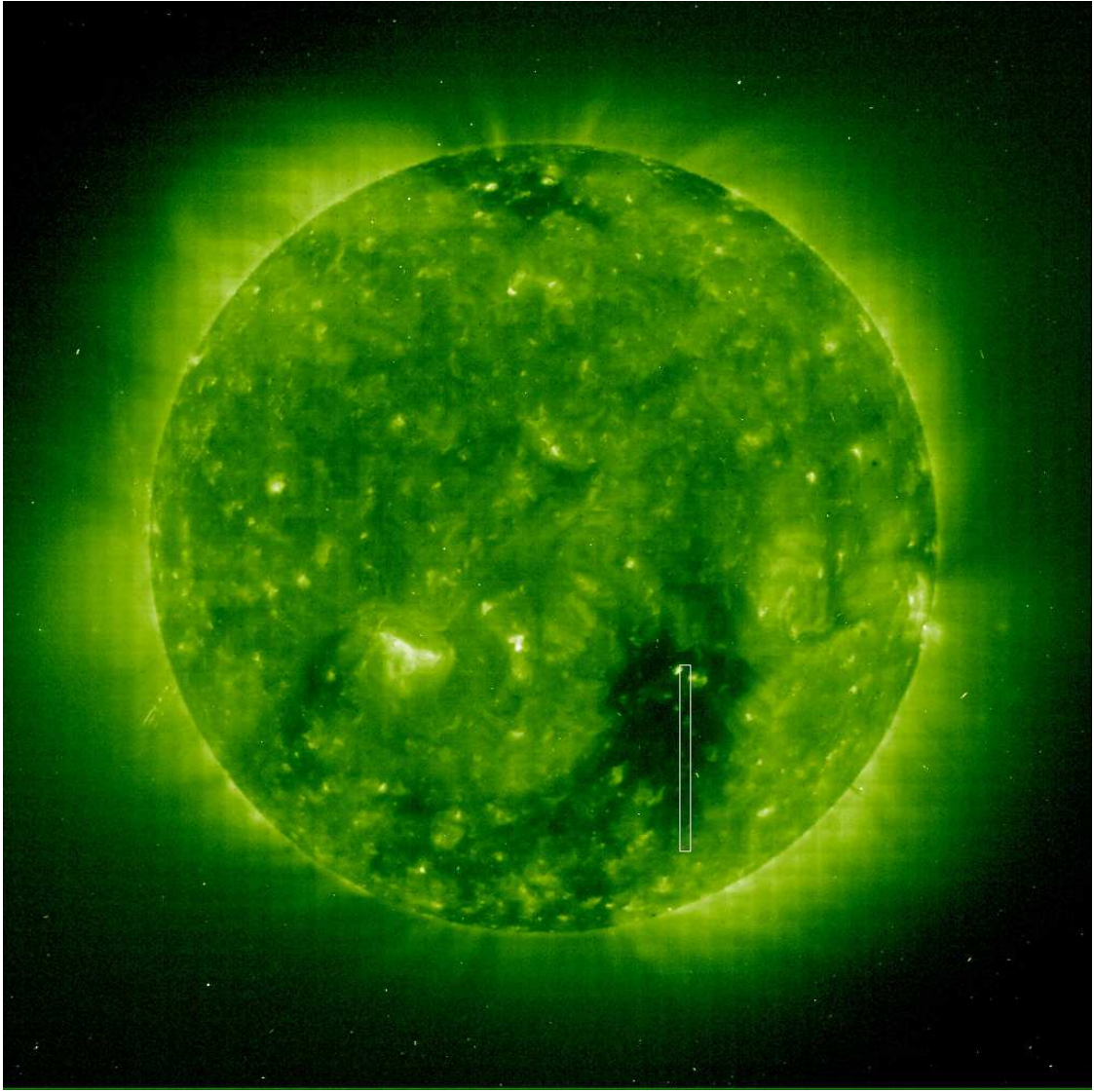}
\caption{\label{fig:context} The box outlines the position of the EIS observation overlayed on a nearly contemporaneous EIT \textit{SOHO} image in the 195~\AA\ band, which consists mainly of Fe~\textsc{xii} emission. 
}
\end{figure}

	We analyzed archival EIS data of a low latitude coronal hole. The observation was performed on 2007 September 26 at 14:17 UT. The 1$^{\prime\prime}$ slit was rastered across 36 positions, each having an exposure time of 90~s. The data come from a 456$^{\prime\prime}$ long portion of the EIS slit. The observation was centered vertically at -537$^{\prime\prime}$ and the slit was rastered from $X=321^{\prime\prime}$ to $349^{\prime\prime}$. Figure~\ref{fig:context} shows the location of this obervation superimposed on an image from the Extreme ultraviolet Imaging Telescope \citep[EIT;][]{Delaboudiniere:SolPhys:1995}. The image was taken at about the same time as the EIS data. The coronal hole we observed appears to be part of the South polar coronal hole.  

	The data were prepared using standard EIS procedures to remove spikes, warm pixels, and CCD dark current, and calibrate the intensity scale. Systematic drifts in the wavelength scale were then corrected using the method developed by \citet{Kamio:SolPhys:2010}. In order to increase statistical accuracy, we binned the $1^{\prime\prime}$ square pixels into bins $9^{\prime\prime}$ horizontally and $19^{\prime\prime}$ vertically. The size was chosen so that each bin would contain the same number of pixels. Thus, the effective exposure time for each bin was about four hours. However this is still rather low compared to the detailed line width studies by \citet{Hahn:ApJ:2012} and \citet{Bemporad:ApJ:2012}, which each used data binned to effective exposure times of over twenty hours.  
	
\begin{figure}
\centering \includegraphics[width=0.5\textwidth]{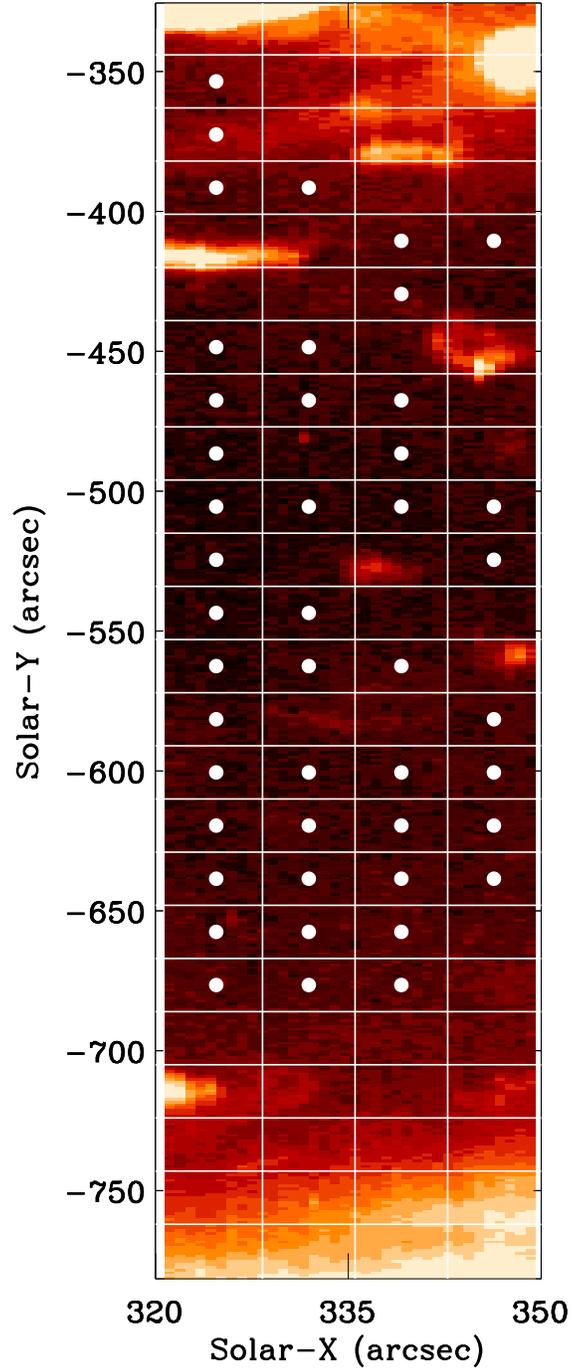}
\caption{\label{fig:selection} Fe~\textsc{xii} 195.12~\AA\ line intensity measured by EIS. The grid shows the binning used in the analysis and the dots indicate those bins that were used in the analysis. These were chosen to avoid bright points and the quiet Sun regions in the bottom and rightmost portions of the observation. 
}
\end{figure}

	Figure~\ref{fig:selection} illustrates the binning as a grid over an unbinned EIS image in the Fe~\textsc{xii}~195~\AA\ line. The observation covers several structures, including the coronal hole, the quiet Sun, and several bright points. As we are interested only in the coronal hole bins we have excluded from our analysis bins that appear to include bright points and quiet Sun material. The remaining bins selected for analysis are marked by the dots in Figure~\ref{fig:selection}. 
	
\subsection{Line Widths} \label{subsec:fit}
	
	Gaussian profiles were fit to each spectral line in order to determine the line width $\Delta \lambda$. In some observations, particularly of active regions, it has been shown that a Doppler-shifted component can cause an apparent broadening of the line profile \citep{Tian:ApJ:2011}. For quiet Sun regions, \citet{McIntosh:ApJ:2009} found blue-shifted components of an Ne~\textsc{viii} having a velocity of $\approx50$~$\mathrm{km\,s^{-1}}$ with a relative intensity of $\approx 5\%$. The shifted components were strongest in the supergranular network. We tested our line profiles for asymmetries by comparing the integrated intensity on each side of the line center determined by a single Gaussian fit and did not find any significant asymmetries. One possible reason for this is that the broad spatial binning covers both network and internetwork regions, which washes out the asymmetry.
	
	Some studies have also shown that line profiles observed on disk have broadened wings and require double Gaussian fits \citep{KjeldsethMoe:ApJ:1977,Wilhelm:SSR:2007}. However, \citet{Peter:AA:2001} measured the relative intensity of the broad and narrow components as a function of line formation temperature and showed that the intensity of the broad component goes to zero at coronal temperatures. We therefore expect single Gaussian fits to be sufficient for our study which focusses on coronal lines. We have checked this by comparing double to single Gaussian fits for some lines, and found no improvement in the fits. Other studies of EIS data from coronal holes have also found that the line profiles are well approximated by a single Gaussian \citep[e.g.,][]{Tian:ApJ:2010}. 
	
	Table~\ref{table:linelist} lists all of the lines used in our analysis; the transitions selected for the line width study are marked with an asterisk. We have avoided using blended lines except for the Mg~\textsc{vi} 270.40~\AA, Fe~\textsc{x} 256.27~\AA, and Fe~\textsc{xii} 195.12~\AA\ doublets, which are self-blends. In order to accurately measure $\Delta \lambda$, we fit these lines with a double Gaussian constrained so that both components have the same width and a fixed separation between the two centroids given by the known wavelengths. 

	The spectral line full width at half maximum $\Delta \lambda_{\mathrm{FWHM}}$ depends on instrumental broadening $\Delta \lambda_{\mathrm{inst}}$, the ion temperature $T_{\mathrm{i}}$, and the non-thermal velocity $v_{\mathrm{nt}}$ \citep{Ultraviolet} as 
\begin{equation}
\Delta \lambda_{\mathrm{FWHM}} = \left[ \Delta\lambda_{\mathrm{inst}}^2 +  
4 \ln(2)\left(\frac{\lambda}{c}\right)^{2}\left(\frac{2k_{\mathrm{B}}T_{\mathrm{i}}}{M} + v_{\mathrm{nt}}^2 \right) \right]^{1/2}, 
\label{eq:width}
\end{equation}	
where $M$ is the ion mass, $k_{\mathrm{B}}$ is the Boltzmann constant, $\lambda$ is the observed wavelength, and $c$ is the speed of light. The instrumental width is known to vary along the length of the slit. \citet{Young:EIS:2011} measured this variation by studying quiet regions near the solar equator. \citet{Hara:ApJ:2011} obtained similar results by comparing the widths of the Fe~\textsc{xiv}~264.78~\AA\ line observed by EIS with ground based observations of the Fe~\textsc{xiv}~5303~\AA\ line. Here, we subtract the instrumental widths using the values from \citet{Young:EIS:2011}. The remaining line width is the sum of the thermal and non-thermal velocities. This width can be expressed in terms of an effective velocity, defined as 
\begin{equation}
v_{\mathrm{eff}}=\sqrt{\left(\frac{2k_{\mathrm{B}}T_{\mathrm{i}}}{M} + v_{\mathrm{nt}}^2 \right)}. 
\label{eq:veffdefine}
\end{equation}

\subsection{Magnetic Field}\label{subsec:mag}
	
	A PFSS model was used to trace the magnetic field lines in the coronal hole \citep{Schatten:SolPhys:1969, Wang:ApJ:1992, Schrijver:SolPhys:2003}. The model assumes that there are no currents between $r=R_{\sun}$ and the source surface at $r=R_{s}$. The radius of the source surface is arbitrary, but a value of $R_{s}=2.5$~$R_{\sun}$ has usually been found to give the best agreement with observations \citep{Hoeksema:JGR:1983}. In the region between $R_{\sun}$ and $R_{\mathrm{s}}$ the magnetic field $\mathbf{B}$ can be determined by solving the Laplace equation. The lower boundary condition is to require that the solution at $r=R_{\sun}$ matches measured solar photospheric magnetogram data. The boundary condition at $r=R_{s}$ is that the field lines become radial. 
	
	Although the PFSS model makes significant simplifications, it generally does a good job of reproducing the large scale structure of the solar magnetic field, particularly for coronal holes \citep{Wang:ApJ:1992, Neugebauer:JGR:1998, Schrijver:SolPhys:2003, Riley:ApJ:2006}. This has been tested by tracing the open field lines, defined as those that reach $R_{s}$, back to the solar surface and comparing these to the observed coronal hole boundaries. The coronal hole boundaries predicted by PFSS and full MHD models are in good agreement with observations. This suggests that PFSS models provide a reasonable approximation for coronal hole magnetic fields. 
		
	Here, we used the model implemented by the PFSS package available in \textit{solarsoft}\footnote{www.lmsal.com/solarsoft} and described in detail by \citet{Schrijver:SolPhys:2003}. The data for this model is available at six hour intervals. We chose the one at 2007-09-26 12:04, which was closest in time to our observation. An inspection of PFSS results from before and after this time showed that the magnetic field lines in the coronal hole appeared to be stable during the observation. In the analysis we corrected for the very small shift in viewing angle due to the rotation of the Sun between time of the PFSS model and the observation time. 
	
	In order to find the angle between the magnetic field line and the line of sight for each spatial bin in our data we used the PFSS package to trace the magnetic field line passing through the center of each bin at a lower height $R_{1}$.  Then we found the coordinates of each of these field lines at a larger height $R_{2}$. The unit vector pointing between these two coordinates gives the average magnetic field direction $\mathbf{\hat{b}}$ between $R_{1}$ and $R_{2}$. We label the direction outward from the Sun parallel to the line of sight as the $\mathbf{\hat{z}}$ direction. The angle between the magnetic field and the line of sight is then $\alpha = \cos^{-1}(\mathbf{\hat{b}} \cdot \mathbf{\hat{z}} )$. 
	
\begin{figure}
\centering \includegraphics[width=0.9\textwidth]{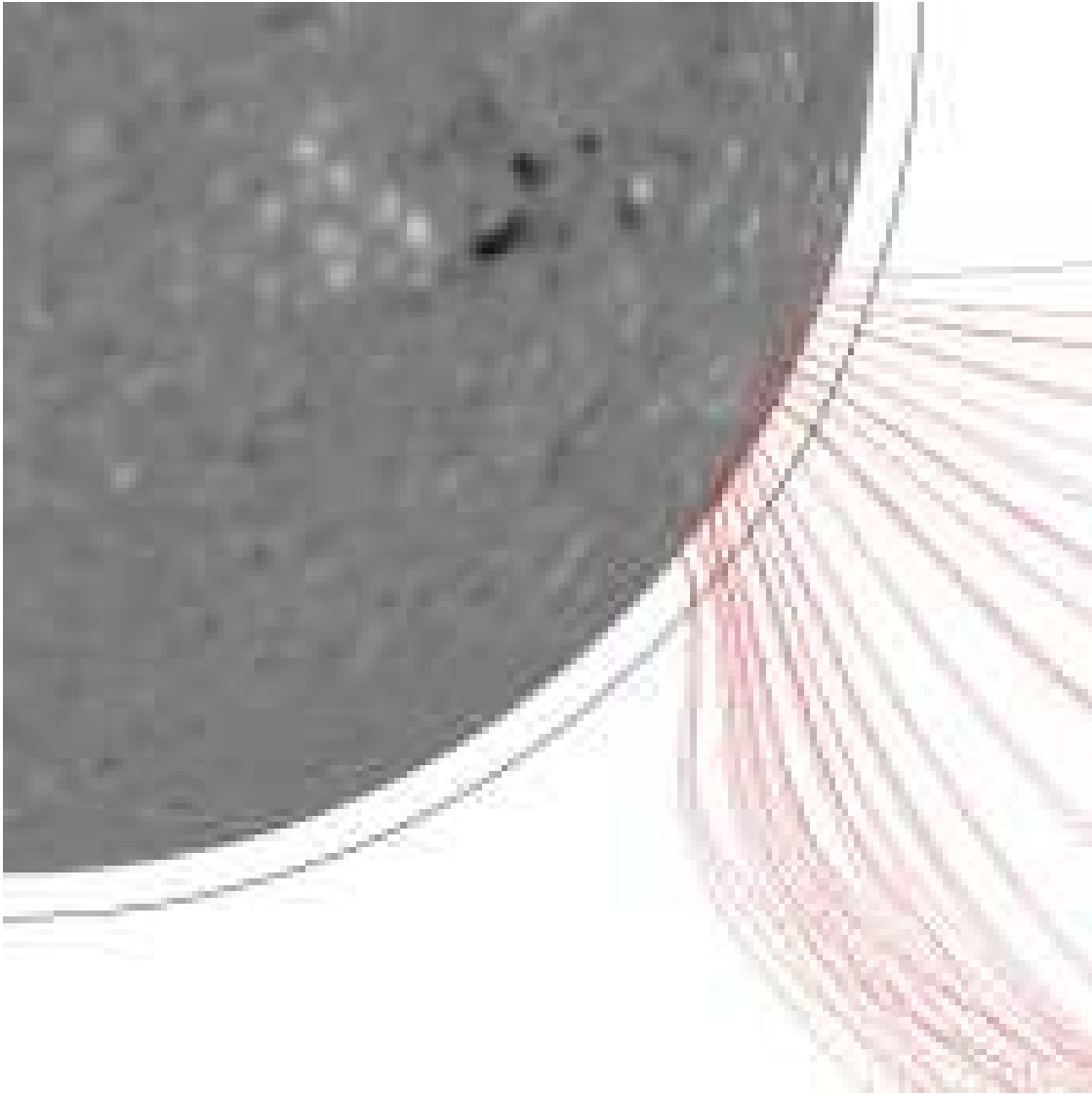}
\caption{\label{fig:pfss} PFSS model tracing the magnetic field lines that pass through the bins used in the analysis (Figure~\ref{fig:selection}). The image has been rotated 50$^{\circ}$ about the North-South axis in order to emphasize the changing angle of the field lines with respect to the line of sight. The dark curve indicates the surface at 1.05~$R_{\sun}$. The grayscale on the solar disk shows the magnetogram data used for the model with white and black indicating positive and negative polarity, respectively \citep{Schrijver:SolPhys:2003}.
}
\end{figure}

	Figure~\ref{fig:pfss} shows the field lines passing through the bin centers. The figure shows that $\alpha$ depends on the radii over which the angle is calculated. The field lines become straighter, and $\alpha$ becomes more constant, for larger heights. Here, we chose $R_{1}=1.02$~$R_{\sun}$ and $R_{2} = 1.05$~$R_{\sun}$ for the analysis. These heights are based on our density and temperature measurements, which are consistent with observing at a height of about 1.02~$R_{\sun}$ with an emission measure scale height of $\approx 0.03$~$R_{\sun}$ (see Section~\ref{subsec:dentemp}). We have also tested our results for line widths and Doppler shifts using height ranges of $1.02$ -- $1.10$~$R_{\sun}$ and $1.01$ -- $1.05$~$R_{\sun}$. The results for the 1.02 -- 1.10~$R_{\sun}$ range compared to 1.02 -- 1.05~$R_{\sun}$ were nearly the same. For 1.01 -- 1.05~$R_{\sun}$ there were larger differences compared to 1.02 -- 1.05~$R_{\sun}$, but the changes were still generally within the $1\sigma$ uncertainties. 

	There are also small variations of the angle $\alpha$ in the plane of the observation with height. Due to the large spatial binning $\alpha$ varies from the nominal value determined at the bin center by $\pm 3^{\circ}$ on average. 
	
	Figure~\ref{fig:pfss} also demonstrates that a simpler assumption of a radial magnetic field would not be accurate. The field lines are radial only far from the boundaries of the coronal hole. 

\subsection{Anisotropy}\label{subsec:anis}
	
	The line width reflects the broadening due to $v_{\mathrm{eff}}$ along the line of sight. The velocity of any particular ion in the observation can be described by $\mathbf{v} = v_{\parallel} \mathbf{\hat{b}} + v_{\perp}\mathbf{\hat{q}}$ where $\mathbf{\hat{b}}$ is the unit vector along the magnetic field and $\mathbf{\hat{q}} \perp \mathbf{\hat{b}}$. Thus, the projection of $\mathbf{v}$ along the line of sight $\mathbf{\hat{z}}$ is $v_{\mathrm{LOS}} = \mathbf{v} \cdot \mathbf{\hat{z}} = v_{\parallel}\cos{\alpha} + v_{\perp}\sin{\alpha}$. 
We expect $v_{\parallel}$ and $v_{\perp}$ to be normally distributed with variances $v_{\mathrm{eff},\parallel}^2$ and $v_{\mathrm{eff},\perp}^2$. Since $\mathbf{v}\cdot \mathbf{\hat{z}}$ is a linear combination of $v_{\parallel}$ and $v_{\perp}$ the variance along the line of sight $v_{\mathrm{eff}}^2$ comes from the convolution of the two Gaussians, which leads to: 
\begin{equation}
v_{\mathrm{eff}}^2 = v_{\mathrm{eff},\parallel}^{2} \cos^{2}(\alpha) + v_{\mathrm{eff},\perp}^2 \sin^{2}(\alpha). 
\label{eq:fitani}
\end{equation}
This is the quantity that we measure from the line width. 
	
\begin{figure}
\centering \includegraphics[width=0.9\textwidth]{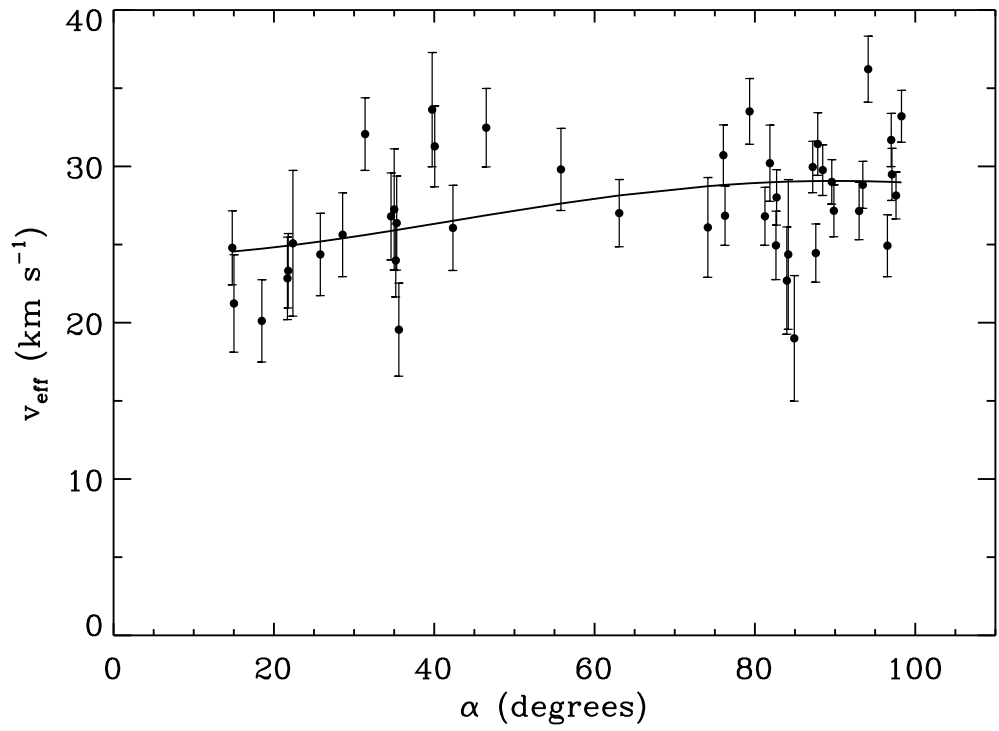}
\caption{\label{fig:anisfit} Effective velocity $v_{\mathrm{eff}}$ versus the angle $\alpha$ between the line of sight and the magnetic field for the Fe~\textsc{xi} lines. The solid line shows the fit to equation~(\ref{eq:fitani}).
}
\end{figure}

	By fitting $v_{\mathrm{eff}}$ versus $\alpha$ we can extract the components $v_{\mathrm{eff},\parallel}$ and $v_{\mathrm{eff},\perp}$ from the measured $v_{\mathrm{eff}}$. A least squares fit was performed separately for each ion species. An example of a fit for the Fe~\textsc{xi} lines is shown in Figure~\ref{fig:anisfit}. As we discuss later, there are some indications that our coarse spatial binning covers different structures within the coronal hole within each bin. If this is the case then a longer observation might remove some of the scatter in the plot by permitting smaller bin sizes. Table~\ref{table:components} summarizes the results for all the ions. The inferred $v_{\mathrm{eff},\parallel}$ and $v_{\mathrm{eff},\perp}$ include both thermal and non-thermal broadening in the same way as described by equation~\ref{eq:veffdefine}. The evaluation of these contributions will be discussed in more detail in Sections \ref{subsec:perp} and \ref{subsec:par}. 

\section{Results and Discussion}\label{sec:res}

\subsection{Density and temperature}\label{subsec:dentemp}

	The electron density $n_{\mathrm{e}}$ and  temperature $T_{\mathrm{e}}$ are useful for interpreting the line width results. The density averaged over the selected bins was determined from several line intensity ratios. From the Fe~\textsc{viii} 185.21~\AA/186.60~\AA\ intensity ratio we inferred an average density of $n_{\mathrm{e}}=1.0 \pm 0.2 \times 10^{8}$~cm$^{-3}$; the Fe~\textsc{ix} 189.94~\AA/188.49~\AA\ intensity ratio gave $n_{\mathrm{e}}=2.1 \pm 0.7 \times 10^{8}$~cm$^{-3}$; and the Fe~\textsc{xiii} 203.8~\AA/202.04~\AA\ ratio implied $n_{\mathrm{e}}=2.6 \pm 1.3 \times 10^{8}$~cm$^{-3}$. Here and throughout all uncertainties are given at an estimated 1$\sigma$ level. Note that the Fe~\textsc{xiii} 203.8~\AA\ line is a blend of four Fe~\textsc{xiii} lines, all of which were included in our analysis. The most important lines are the ones at 203.796~\AA\ and 203.827~\AA, but for $n_{\mathrm{e}} \lesssim 5 \times 10^{8}$~cm$^{-3}$ contributions from the lines at 203.772~\AA\ and 203.835~\AA\ are also important.
	
	These values are within a factor of a few, which is interesting since the Fe~\textsc{viii}, \textsc{ix} and \textsc{xiii} ions are formed over a range of temperatures. Both the transition region and the corona lie along the line of sight, so the observation looks into a temperature gradient with $T_{\mathrm{e}}$ increasing with height. In contrast, the density should be decreasing with height. Here we find that the densities roughly agree, or possibly follow a temperature dependence opposite to what is expected, although the uncertainties are too large to say so definitively. This agreement suggests that the ions are formed in the same volume. However, another possibility is that there are unresolved hotter denser structures that skew the density measurements.
	
	For an on-disk observation there will be material at a range of different temperatures along the line of sight. In order to measure $T_{\mathrm{e}}$ we performed a differential emission measure (DEM) analysis. The DEM $\phi(T_{\mathrm{e}})$ describes the amount of material along the line of sight as a function of $T_{\mathrm{e}}$. In terms of $\phi(T_{\mathrm{e}})$, the intensity of an emission line emitted by a transition from level $j$ to level $i$ is given by 
\begin{equation}
I_{ji} = \frac{1}{4\pi}\int{ G(T_{\mathrm{e}})\phi(T_{\mathrm{e}}) dT_{\mathrm{e}}},
\label{eq:intensity}
\end{equation}
where $G(T_{\mathrm{e}})$ is the contribution function and describes the level populations, ionization balance, elemental abundance, and radiative decay rates. These data are available in the CHIANTI atomic database \citep{Dere:AA:1997,Landi:ApJ:2012}. Since we started this work the CHIANTI database has been updated from Version 7 to Version 7.1, but these updates have little effect on the plasma DEM and do not affect the analysis of line widths and Doppler shifts \citep{Landi:Private}. Given $G(T_{\mathrm{e}})$ and a set of measured line intensities $I_{ji}$, it is possible to invert equation~(\ref{eq:intensity}) to find $\phi(T_{\mathrm{e}})$. 

	We calculated the DEM using the technique described in \citet{Landi:AA:1997}. The lines used for this analysis are listed in Table~\ref{table:linelist} and include some ions formed at lower temperatures than those considered in the line width analysis. For the inversion it was necessary to set the behavior of $\phi(T_{\mathrm{e}})$ at both the high and low temperature ends. As was done in \citet{Hahn:ApJ:2011}, at high temperatures we assumed that $\phi(T_{\mathrm{e}})=0$ at $\log T_{\mathrm{e}}=8$ (here and throughout all temperatures are reported in Kelvin). This is a reasonable condition since there should not be any material at higher temperatures in the observation. Due to the presence of the transition region and chromosphere, $\phi(T_{\mathrm{e}})$ will not be zero at low temperatures and it was not possible to anticipate the low temperature behavior of the DEM prior to performing the analysis. For the DEM reported here we have set $\phi(T_{\mathrm{e}})$ at low temperatures to a constant value determined by the He~\textsc{ii} line, which is the coolest line in our data. We also tested a variety of other conditions and found that for $\log T_{\mathrm{e}} \gtrsim 5.6$, the DEM does not vary significantly. This is because the contribution function for the coronal lines goes to zero at low temperatures, so that temperature range contributes nothing to the coronal line intensity integrals. As an additional check we also performed the DEM analysis using the regularized inversion method of \citet{Hannah:AA:2012}. Both DEM inversion methods agree to within their respective uncertainties for $\log T_{\mathrm{e}} \gtrsim 5.6$. Thus, despite the poor constraints at low temperatures the DEM appears to be valid for coronal temperatures. 

\begin{figure}
\centering \includegraphics[width=0.9\textwidth]{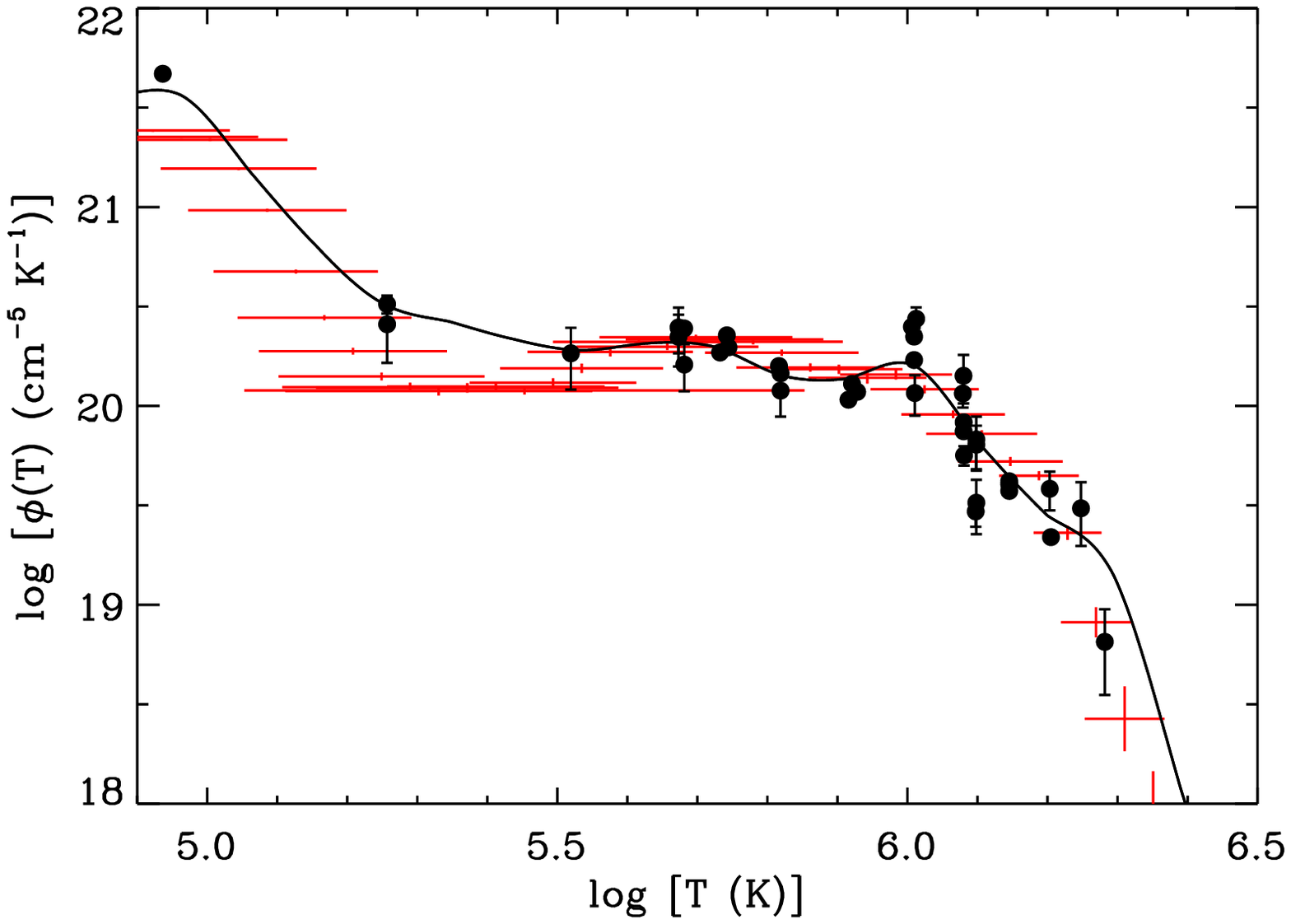}
\caption{\label{fig:DEM} Representative DEM for the coronal hole. The solid line shows the results from the \citet{Landi:AA:1997} inversion method and the crosses show the results from the \citet{Hannah:AA:2012} method. The filled circles show the points on the DEM determined from the individual measured line intensities and their scatter gives an estimate of the uncertainty in the \citet{Landi:AA:1997} inferred DEM. On the temperature scale, these points correspond to $\log T_{t}$ for each line (see text). The size of the various error bars indicates the $1\sigma$ uncertainties. 
}
\end{figure}

	We calculated the DEM for various positions within the observation and found that the DEMs from all the selected coronal hole bins were similar. Figure~\ref{fig:DEM} shows the DEM calculated for a typical position in the coronal hole, $X=325^{\prime\prime}$ and $Y=-525^{\prime\prime}$. In the figure the solid line indicates the DEM calculated using the \citet{Landi:AA:1997} method and the crosses represent the DEM from the \citet{Hannah:AA:2012} method. The DEM shows that the observed emission comes from a broad range of temperatures, though there is a peak at $\log T_{\mathrm{e}} \approx 6$. The dots in the figure correspond to the various lines used in the DEM analysis. The error bars on these points are from the uncertainty of intensity only, and neglect possible additional errors from atomic data and unknown blends. Their position on the temperature scale corresponds to the DEM-averaged temperature at which each line is emitted: 
\begin{equation}
\log T_{t} = \frac{\int G_{ji}(T_{\mathrm{e}})\phi(T_{\mathrm{e}}) \log T_{\mathrm{e}} dT_{\mathrm{e}}}
	{\int G_{ji}(T_{\mathrm{e}}) \phi(T_{\mathrm{e}}) dT_{\mathrm{e}}}.
\label{eq:avgtemp}
\end{equation}
To determine the $T_{t}$, we used the \citet{Landi:AA:1997} method DEM results. Since the temperature dependence of $G_{ji}$ is mainly due to the ionization balance, the effective formation temperature $T_{t}$ is about the same for all lines from a given ion species. Table~\ref{table:components} lists $T_{t}$ for the ions used in the line width analysis.

	The DEM has a small peak at about $\log T_{\mathrm{e}} = 6.0$ and drops off towards higher temperatures, but at lower temperatures $\phi$ does not drop due to presence of the transition region along the line of sight. The extension of the DEM up to $\log T_{e} \approx 6.2$ may be due to the limited temperature resolving power of the DEM inversion. \citet{Landi:AA:2012} studied the ability of a DEM technique to measure isothermal plasmas and found that the method could not resolve two isothermal components if they were separated by less than about $\Delta \log T_{\mathrm{e}} = 0.20$. \citet{Hahn:ApJ:2011} use the same DEM technique as here to study an off-limb coronal hole observation, and found that the full width half maximum for the peak in $\phi(T_{\mathrm{e}})$ was about $\log T_{\mathrm{e}} =0.15$. This raises the possibility that the lines from higher charge states, such as Fe~\textsc{xii} and \textsc{xiii}, include contributions from hotter structures.	Previously, \citet{Hahn:ApJ:2011} measured the DEM of a polar coronal hole and suggested that the high temperature tail could be due to quiet Sun coronal material intervening along the line of sight.  However, this explanation does not apply to the present observation where the PFSS model shows that the closed field lines are bent away from the coronal hole and do not intersect the line of sight. In this case a more likely explanation is that the large spatial binning sums over temperature variation due to small scale structures within the coronal hole. 
	
	 Because the intensity is proportional to $n_{\mathrm{e}}^2$, the observed emission will be dominated by the points along the line of sight with the greatest density. The measured $\log T_{\mathrm{e}} \sim 6$ implies a density scale height of $\approx 0.06$~$R_{\sun}$. Since the scale height for $n_{\mathrm{e}}^2$ is half the scale height for $n_{\mathrm{e}}$, we infer that the measured intensities come primarily from a height range of $\approx 0.03$~$R_{\sun}$, assuming a constant temperature. Our density measurements can be used to specify the height in the corona where the emission is produced. Off-limb density measurements by \citet{Hahn:ApJ:2010} and \citet{Bemporad:ApJ:2012} showed that the measured density of $n_{\mathrm{e}}\approx 2\times10^{8}$~$\mathrm{cm}^{-3}$ corresponds to heights of about 1.02~$R_{\sun}$ to 1.05~$R_{\sun}$. Thus, the observed emission appears to come from approximately these heights.

\subsection{Perpendicular Broadening}\label{subsec:perp}

	The component $v_{\mathrm{eff},\perp}$ perpendicular to the magnetic field depends on $T_{\mathrm{i},\perp}$ and $v_{\mathrm{nt},\perp}$ through a relation analogous to equation~(\ref{eq:veffdefine}), namely
\begin{equation}
v_{\mathrm{eff}, \perp}=\sqrt{\left(\frac{2k_{\mathrm{B}}T_{\mathrm{i},\perp}}{M} + v_{\mathrm{nt},\perp}^2 \right)}. 
\label{eq:veffperpdefine}
\end{equation}	
In coronal holes $T_{\mathrm{i},\perp}$ has been observed to be greater than $T_{\mathrm{e}}$ and varies with the ion charge to mass ratio \citep{Esser:ApJ:1999,Landi:ApJ:2009,Hahn:ApJ:2010}. The perpendicular heating may be due to such things as ion cyclotron resonance with high frequency waves \citep{Cranmer:SSR:2002, Hollweg:JAA:2008}, or stochastic heating by turbulence \citep{Voitenko:ApJ:2004, Chandran:ApJ:2010}. The non-thermal velocity perpendicular to the magnetic field is thought to be proportional to the amplitude of Alfv\'en waves \citep{Doyle:SolPhys:1998,Banerjee:AA:1998,Banerjee:SSR:2011}. 

	Upper and lower bounds on $T_{\mathrm{i},\perp}$ and an upper bound for $v_{\mathrm{nt},\perp}$ were found from $v_{\mathrm{eff},\perp}$ using a method based on that of \citet{Tu:ApJ:1998}. Their approach assumes that all the emission comes from the same volume so that the fluid motion from the waves, seen as $v_{\mathrm{nt},\perp}$, is the same for all the ions. In our observation the volumes are large and could encompass different structures, which might make this assumption invalid.  However, as discussed below, our results do appear to be consistent with $v_{\mathrm{nt},\perp}$ the same for all the ions. 
	
	The upper bound for $T_{\mathrm{i},\perp}$ was determined by setting $v_{\mathrm{nt},\perp}=0$. To find the lower bound \citet{Tu:ApJ:1998} assumed that for the narrowest line the width was entirely due to non-thermal broadening and applied that value of $v_{\mathrm{nt},\perp}$ to determine $T_{\mathrm{i},\perp}$ for the remaining lines.	We found that assumption leads to a lower bound on $T_{\mathrm{i},\perp} > T_{\mathrm{e}}$ for most lines. Since we expect that the ions will either be in equilibrium with the electrons or be heated, we followed \citet{Hahn:ApJ:2010} and derived a tighter constraint on the lower bound of $T_{\mathrm{i},\perp}$ by assuming $T_{\mathrm{i},\perp} \geq T_{\mathrm{e}}$. The maximum $v_{\mathrm{nt},\perp}$ is given by the minimum width after subtracting this lower bound of $T_{\mathrm{e}}$ from $v_{\mathrm{eff},\perp}$. For this analysis we took $\log T_{\mathrm{e}} = 5.67$, since this was the lowest value of $\log T_{t}$ for these ions (Table~\ref{table:components}).
			
\begin{figure}
\centering \includegraphics[width=0.9\textwidth]{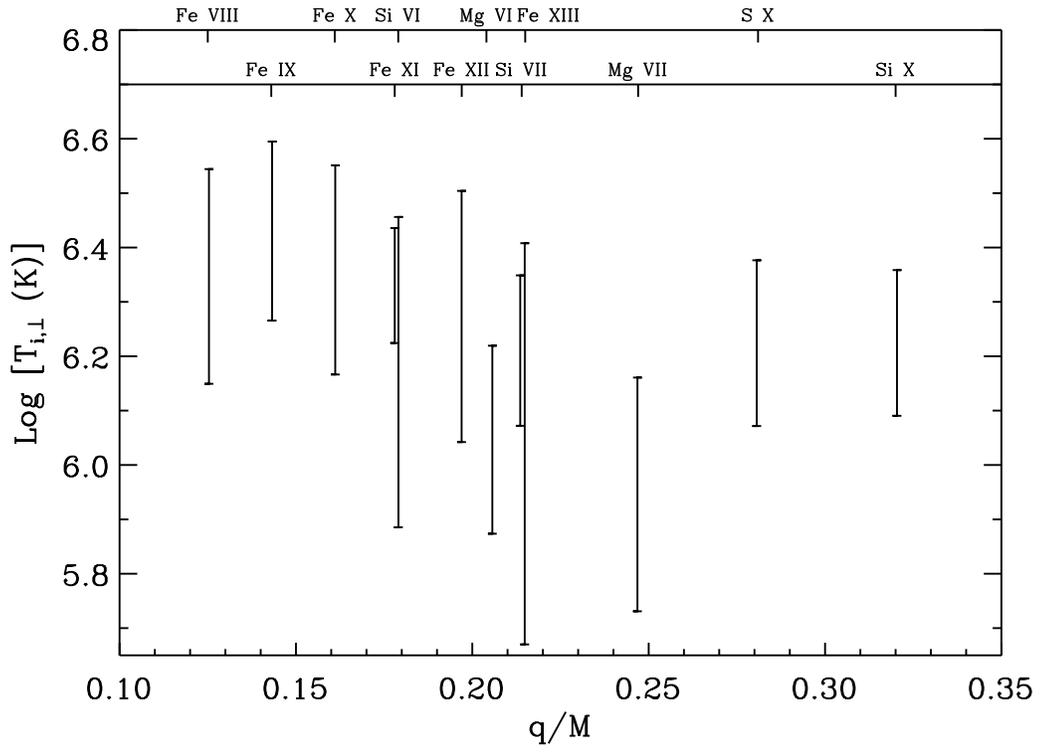}
\caption{\label{fig:Tiperp} Upper and lower bounds for $\log T_{\mathrm{i},\perp}$ as a function of charge to mass ratio $q/M$. The top axis labels the corresponding ion species. 
}
\end{figure}

	Figure~\ref{fig:Tiperp} shows the upper and lower bounds for $T_{\mathrm{i},\perp}$ as a function of charge to mass ratio $q/M$. 	For low $q/M$ ions, even the lower bounds for $T_{\mathrm{i},\perp}$ are greater than $T_{\mathrm{t}}$, which we take as a measure of $T_{\mathrm{e}}$ for each ion. The inferred $T_{\mathrm{i},\perp}$ decreases for $q/M \gtrsim 0.16$. These results can be compared to similar measurements by \citet{Landi:ApJ:2009} and \citet{Hahn:ApJ:2010} from off-disk spectra. The shape of $T_{\mathrm{i},\perp}$ versus $q/M$ is similar, with a high temperature at low $q/M$ that drops at higher $q/M$. However, in the present observation the ion temperature for the lowest $q/M$ ions was smaller than that found in off-limb observations where $\log T_{\mathrm{i},\perp}$ was $\gtrsim 6.5$. 
	
	A possible explanation for these differences is that here, the data come from a lower height in the corona. The ions are expected to be cooled through collisions with the protons, which are expected to be cooler than the ions. The density here is about $2\times10^{8}$~cm$^{-3}$, compared to $8\times10^{7}$~cm$^{-3}$ at the lowest height in the off-limb observation of \citet{Hahn:ApJ:2010}. Thus, collisional cooling is more important and this may make $T_{\mathrm{i},\perp}$ smaller here. 
	
	The coarse spatial binning of our data introduces a possible systematic error through the assumption that $v_{\mathrm{nt},\perp}$ is the same for all the ions because different structures might not have the same $v_{\mathrm{nt},\perp}$. We can test the validity of this assumption by checking the consistency of $T_{\mathrm{i},\perp}$ for ions formed at different $T_{t}$. For example, Si~\textsc{vi} at $\log T_{t} = 5.67$ and Fe~\textsc{xi} at $\log T_{t} = 6.08$ have nearly the same $q/M$ and their $T_{\mathrm{i},\perp}$ are in reasonable agreement. Another example is the similar level of agreement in $T_{\mathrm{i},\perp}$ found between Si~\textsc{vi} and Fe~\textsc{xiii}. The consistency of the results suggests that our assumptions in the analysis were justified. However, we should note that the possible presence of different structures in the data is not ruled out. It is possible that there are multiple structures, but they happen to have the same $v_{\mathrm{nt},\perp}$, or that our measurements are insensitive to small inhomogeneities. 
	
	The above analysis for $T_{\mathrm{i},\perp}$ yielded an upper bound for $v_{\mathrm{nt},\perp}$ of 24.9~$\mathrm{km\,s^{-1}}$. For undamped Alfv\'en waves on open field lines we expect $v_{\mathrm{nt}, \perp} \propto n_{\mathrm{e}}^{-1/4}$ \citep{Moran:AA:2001}. Off-limb observations have shown that $v_{\mathrm{nt},\perp}$ follows this dependence, which supports the interpretation of $v_{\mathrm{nt},\perp}$ as Alfv\'en waves. Those observations have also shown that the waves are undamped below $\approx 1.15$~$R_{\sun}$ \citep{Doyle:SolPhys:1998, Banerjee:AA:1998, Hahn:ApJ:2012, Bemporad:ApJ:2012}. Thus, we can use this proportionality to scale our measured $v_{\mathrm{nt},\perp}$ for comparison with off-limb observations. For $n_{\mathrm{e}}\approx2\times10^{8}$~cm$^{-3}$ here and a typical value of $8\times10^{7}$~cm$^{-3}$ at $1.05$~$R_{\sun}$ in off-limb observations, the scaling factor is $\approx 1.26$ and the expected upper bound for $v_{\mathrm{nt},\perp}$ at $1.05$~$R_{\sun}$ is 31.4~$\mathrm{km\,s^{-1}}$. This is close to the values of $v_{\mathrm{nt},\perp} \leq 15$ -- 30~$\mathrm{km\,s^{-1}}$ found by \citet{Hahn:ApJ:2010}. \citet{Landi:ApJ:2009} performed a similar analysis, but with the condition $T_{\mathrm{i}, \perp} \geq 0$ instead of $T_{\mathrm{i},\perp} \geq T_{\mathrm{e}}$, and found $v_{\mathrm{nt},\perp}\leq30$ -- 35~$\mathrm{km\,s^{-1}}$. Using the same conditions as theirs for $T_{\mathrm{i},\perp}$ to reanalyze our data, we find $v_{\mathrm{nt},\perp} \leq 28.3$~$\mathrm{km\,s^{-1}}$. This scales to $35.7$~$\mathrm{km\,s^{-1}}$ at 1.05~$R_{\sun}$, in reasonable agreement with \citet{Landi:ApJ:2009}. Thus, the values for $v_{\mathrm{nt},\perp}$ found on-disk appear consistent with the off-limb observations. 

\subsection{Parallel Broadening} \label{subsec:par}

The relation among $v_{\mathrm{eff},\parallel}$, $T_{\mathrm{i},\parallel}$, and $v_{\mathrm{nt},\parallel}$ is analogous to equation~(\ref{eq:veffperpdefine}). The thermal and non-thermal contributions in this direction can be estimated using the fact that both stochastic or ion cyclotron resonant heating are predicted to be weak in the parallel direction \citep{Cranmer:ApJ:1999,Chandran:ApJ:2010}, and so $T_{\mathrm{i},\parallel}$ is set by collisions with protons and electrons. 

In the low corona electron-ion collisions are common. The Spitzer electron-ion temperature equilibration time is  \citep{Spitzer:Book:1962,Bochsler:ARAA:2007}
\begin{equation}
\tau_{\mathrm{eq,e}} = \frac{3Mm_{\mathrm{e}}}{32\pi^{1/2}n_{\mathrm{e}}Z_{\mathrm{i}}^{2}e^{4}\ln{\Lambda}}
\left(\frac{2k_{\mathrm{B}}T_{\mathrm{i}}}{M} + \frac{2k_{\mathrm{B}}T_{\mathrm{e}}}{m_{\mathrm{e}}}\right)^{3/2},
\label{eq:spitzer}
\end{equation}
where $Z_{\mathrm{i}}$ is the ion charge, $e$ is the elementary charge, $m_{\mathrm{e}}$ is the electron mass, and $\ln{\Lambda}$ is the Coulomb logarithm, which is $\sim 20$ for the corona. To estimate $\tau_{\mathrm{eq,e}}$ we can take $n_{\mathrm{e}}= 2\times10^{8}$~cm$^{-3}$, and $T_{\mathrm{i}}=T_{\mathrm{e}}=10^{6}$~K. The electron-ion temperature equilibration time is then $\tau_{\mathrm{eq,e}} \approx 30$~s. Because of their more similar masses proton-ion collisions result in a more efficient energy exchange so that the proton-ion equilibration time $\tau_{\mathrm{eq,p}}$ is about $\sqrt{m_{\mathrm{e}}/m_{\mathrm{p}}}$ shorter than $\tau_{\mathrm{eq,e}}$, or about 0.7~s here. This can be compared to the outflow timescale through the observed height. \citet{Tian:ApJ:2010} has found the outflow speed in on-disk coronal hole measurements to be $\lesssim 10$~$\mathrm{km\,s^{-1}}$. We estimate the height range of the observation to be $ > 0.01$~$R_{\sun}$. Thus, the outflow timescale is $\tau_{\mathrm{out}} \gtrsim 700$~s, which is much longer than the equilibration time. This argument is supported by computational models for ion heating, which are consistent with $T_{\mathrm{i},\parallel}=T_{\mathrm{p}}$ and isotropic $T_{\mathrm{p}}$ at such low heights in the corona of $\lesssim 1.3$~$R_{\sun}$ \citep{Cranmer:ApJ:1999,Landi:ApJ:2009}. 

	Since parallel heating is predicted to be weak, we expect all the ions should be in equilibrium with the protons and all the ions should have the same $T_{\mathrm{i},\parallel} = T_{\mathrm{p}}$. Additionally, if the emission comes from the same volume then all the ions are subject to the same fluid motions so that $v_{\mathrm{nt}, \parallel}$ is the same for all the ions. To test whether these conditions are consistent with our data we found values for $v_{\mathrm{nt},\parallel}$ and $T_{\mathrm{i},\parallel}$ that match the measured $v_{\mathrm{eff},\parallel}$ for each measured ion species to within $2\sigma$ or better. This test shows that $v_{\mathrm{nt},\parallel} =  6.0$ -- $14.9$~$\mathrm{km\,s^{-1}}$ and $T_{\mathrm{i},\parallel} = 1.5 \times 10^{6}$ -- $2.1\times10^{6}$~K are consistent with the data. Note that, as in the perpendicular case, for the analysis we implicitly assumed that the possible presence of multiple structures within each bin do not affect the analysis. This assumption is justified a posteriori by the consistency of our results. A possible reason that the higher $T_{\mathrm{e}}$ structures do not affect the analysis is that their $v_{\mathrm{nt},\parallel}$ and $T_{\mathrm{i},\parallel}$ may not be too different from those of the cooler structures. This could be caused by having different electron heating, but similar waves and proton and ion heating in the different structures. 
	
	Having established that nearly uniform values for the ion temperature and nonthermal velocity are consistent with the data, we determined the most probable values by performing a least squares fit to the parallel analog of Equation~\ref{eq:veffdefine} using the measured $v_{\mathrm{eff},\parallel}$. In this fit the known ion masses $M$ are the only independent variables. We find that $T_{\mathrm{i},\parallel} = (1.8 \pm 0.2) \times 10^{6}$~K and $v_{\mathrm{nt},\parallel} = 12.6 \pm 2.3$~$\mathrm{km\,s^{-1}}$. 

\begin{figure}
\centering \includegraphics[width=0.9\textwidth]{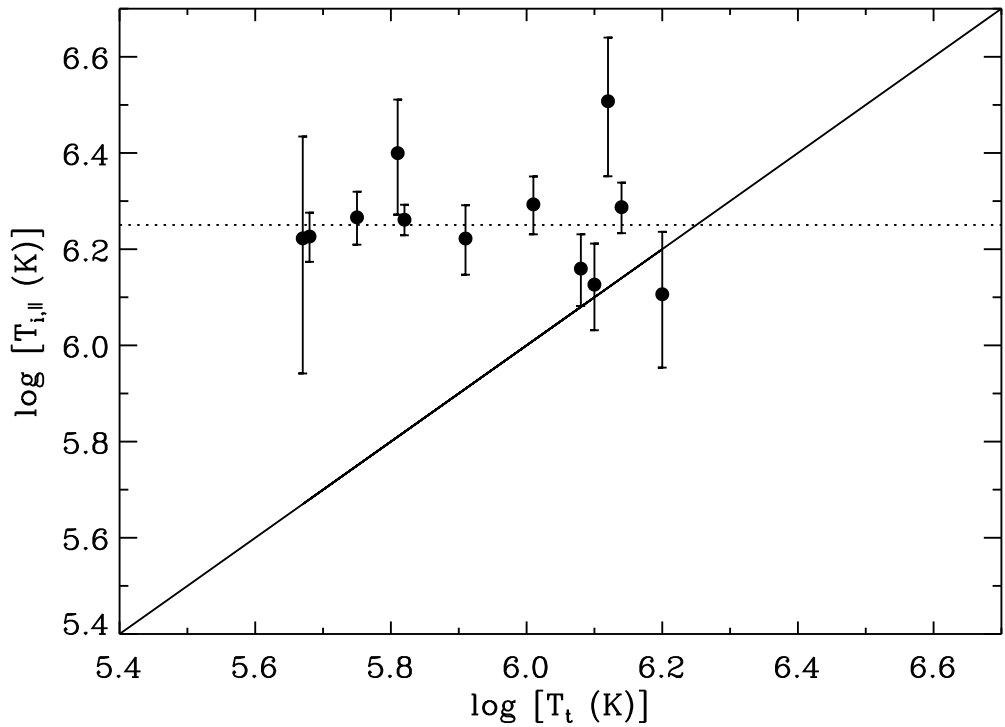}
\caption{\label{fig:Tipar} Parallel ion temperature $\log T_{\mathrm{i},\parallel}$ as a function of the DEM-averaged formation temperature $\log T_{t}$. The filled circles indicate $T_{\mathrm{i},\parallel}$ after setting $v_{\mathrm{nt},\parallel}$ to the fitted value of $v_{\mathrm{nt},\parallel}=12.6 \pm 2.3$~$\mathrm{km\,s^{-1}}$. The dotted line shows the best fit temperature of $T_{\mathrm{i},\parallel} = (1.8 \pm 0.2) \times 10^{6}$~K. The solid line illustrates where $T_{\mathrm{i},\parallel} = T_{t}$.
}
\end{figure}

	Figure~\ref{fig:Tipar} shows $T_{\mathrm{i},\parallel}$ versus $T_{t}$ after setting $v_{\mathrm{nt},\parallel}$ to the fitted value of $12.6$~$\mathrm{km\,s^{-1}}$. The dotted line indicates the fitted $T_{\mathrm{i},\parallel}$ and the solid line illustrates $T_{\mathrm{i},\parallel}=T_{t}$. The plot shows that the inferred $T_{\mathrm{i},\parallel}$ is greater than $T_{t}$ over a wide range of formation temperatures. 
	
	One possible interpretation for $T_{\mathrm{i},\parallel} > T_{t}$ is that the ions reflect a proton temperature that is roughly twice $T_{\mathrm{e}}$. Our $T_{\mathrm{i},\parallel}$ value does fall within the range of estimated proton temperatures in the corona, though there are few measurements of $T_{\mathrm{p}}$ at low heights. \citet{Marsch:AA:2000} have measured $T_{\mathrm{p}} \approx 2 \times 10^{5}$~K at $\approx 1.02$~$R_{\sun}$ based on hydrogen line widths. We have estimated the height of this observation to be within about $0.03$~$R_{\sun}$ above 1.02~$R_{\sun}$. Thus, our data imply a large temperature gradient in this height range. Measurements with UVCS of the Ly$\alpha$ line above 1.3~$R_{\sun}$ have shown $T_{\mathrm{p}} \geq T_{\mathrm{e}}$, with a value of up to $3\times10^{6}$~K  \citep{Esser:ApJ:1999, Antonucci:SolPhys:2000}. Our estimate for $T_{\mathrm{i},\parallel}$ falls between the proton temperature measurements at slightly smaller and somewhat larger heights. This may provide a useful constraint for solar wind models.

	Comparing $T_{\mathrm{i},\parallel}$ to $T_{\mathrm{i},\perp}$, we find that because of the large spread in the lower and upper bounds for $T_{\mathrm{i},\perp}$ the ion temperatures in the parallel and perpendicular directions are similar. This may be due to either strong collisional cooling of the ions, most likely by protons, or weak perpendicular heating at these heights. Only for small $q/M < 0.18$ are the data consistent with $T_{\mathrm{i},\perp} > T_{\mathrm{i},\parallel}$. In contrast, UVCS measurements at 2.2~$R_{\sun}$ have shown $T_{\mathrm{i},\perp} > T_{\mathrm{i},\parallel}$ for O~\textsc{vi} ($q/M = 0.31$). This implies that perpendicular heating of higher $q/M$ ions becomes more efficient at larger heights.

	The parallel non-thermal velocity is thought to be due to slow magnetosonic waves \citep{McClements:SolPhys:1991,Erdelyi:AA:1998}. In addition to broadening the line widths, slow waves are predicted to produce asymmetric line shapes, which we did not find in this observation. \citet{Verwichte:ApJ:2010} has, however, shown that for low amplitude waves with $v_{\mathrm{nt},\parallel} \lesssim 15$~$\mathrm{km\,s^{-1}}$ the asymmetry would be only a few percent. Thus, for our inferred $v_{\mathrm{nt},\parallel}$ of $12.6 \pm 2.3 $~$\mathrm{km\,s^{-1}}$ the asymmetry is predicted to be small, which may explain why we did not detect any. 
	
	 Magnetosonic waves are compressional, and can therefore be observed as time varying line intensity oscillations \citep{Banerjee:SSR:2011}. In polar coronal hole plumes, \citet{DeForest:ApJ:1998} found intensity variations corresponding to density amplitudes of 5\% - 10\% propagating near the sound speed. Similar perturbations in interplume regions have been found by \citet{Banerjee:AA:2001} with amplitudes also of a few percent. These results suggest that the observed oscillations are due to slow magnetosonic waves. \citet{Banerjee:AA:2001} estimated the amplitude  for the observed waves to be $\approx 10$~$\mathrm{km\,s^{-1}}$. This amplitude is consistent with our inferred value for $v_{\mathrm{nt},\parallel}$. 
	 
\subsection{Outflow Velocity}\label{subsec:dopp}

\begin{figure}
\centering \includegraphics[width=0.9\textwidth]{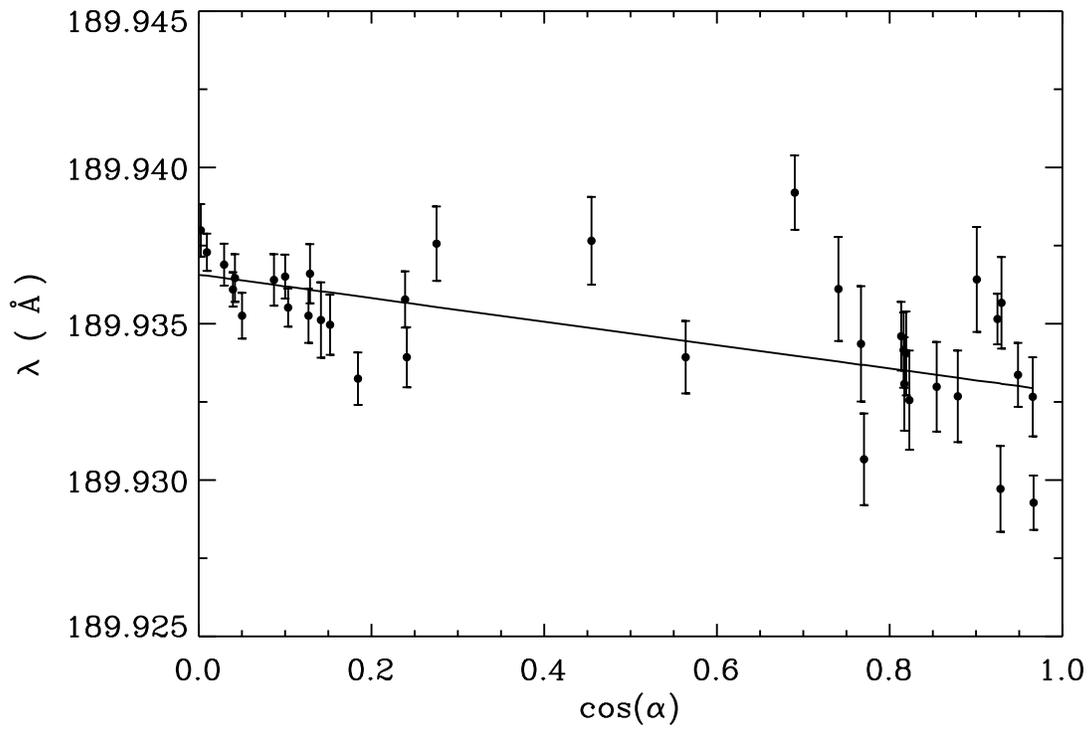}
\caption{\label{fig:dopfit} Linear fit of $\lambda$ versus $\cos(\alpha)$ for the Fe~\textsc{ix} 189.94~\AA\ line. For this particular line the fit finds $\lambda_{0} = 189.9362 \pm 0.0002$~\AA\ and $v_{\parallel} = 5.9 \pm 0.6$~$\mathrm{km\,s^{-1}}$. 
}
\end{figure}

	Combining measured Doppler shifts with the known magnetic field direction provides a new method for determining the outflow velocity. The velocity projection along the line of sight of the bulk flows parallel, $v_{\mathrm{\parallel}}$, and perpendicular, $v_{\mathrm{\perp}}$, to the magnetic field is given by $v_{\mathrm{LOS}} = v_{\parallel}\cos{\alpha} + v_{\perp}\sin{\alpha}$. This line of sight velocity produces a Doppler shift given by $\lambda_{0} - \lambda = \lambda_{0}(v_{\mathrm{LOS}}/c)$, where $\lambda_{0}$ is the rest wavelength. Note that with this definition a positive velocity corresponds to a blueshift. For the corona, the transverse motion of the magnetic field lines is expected to be small, $v_{\perp}\approx0$. Thus, $\lambda_{0}$ and $v_{\parallel}$ can be determined from a linear least squares fit to $\lambda$ versus $\cos{\alpha}$: 
\begin{equation}
\lambda = \lambda_{0} - \lambda_{0}\frac{v_\parallel}{c}\cos{\alpha}. 
\label{eq:doppler}
\end{equation}
An example fit is shown in Figure~\ref{fig:dopfit}. Differential rotation could cause systematic errors in the analysis since $\alpha$ varies with latitude. However, based on the rotation rates measured by \citet{Wohl:AA:2010} we estimate that the rotation velocity differs by less than $0.2$~$\mathrm{km\,s^{-1}}$ over the latitude range in our data. We therefore neglect differential rotation in the analysis. The rest wavelength on the detector $\lambda_{0}$ and its $1\sigma$ uncertainty were found directly from the least squares fit, and these values were used in deriving $v_\parallel$ and its uncertainty from the fitted slope. We found that, on average, the inferred $\lambda_{0}$ agreed with the wavelengths tabulated in CHIANTI \citep{Dere:AA:1997, Landi:ApJ:2012} to within 0.004~\AA.

\begin{figure}
\centering \includegraphics[width=0.9\textwidth]{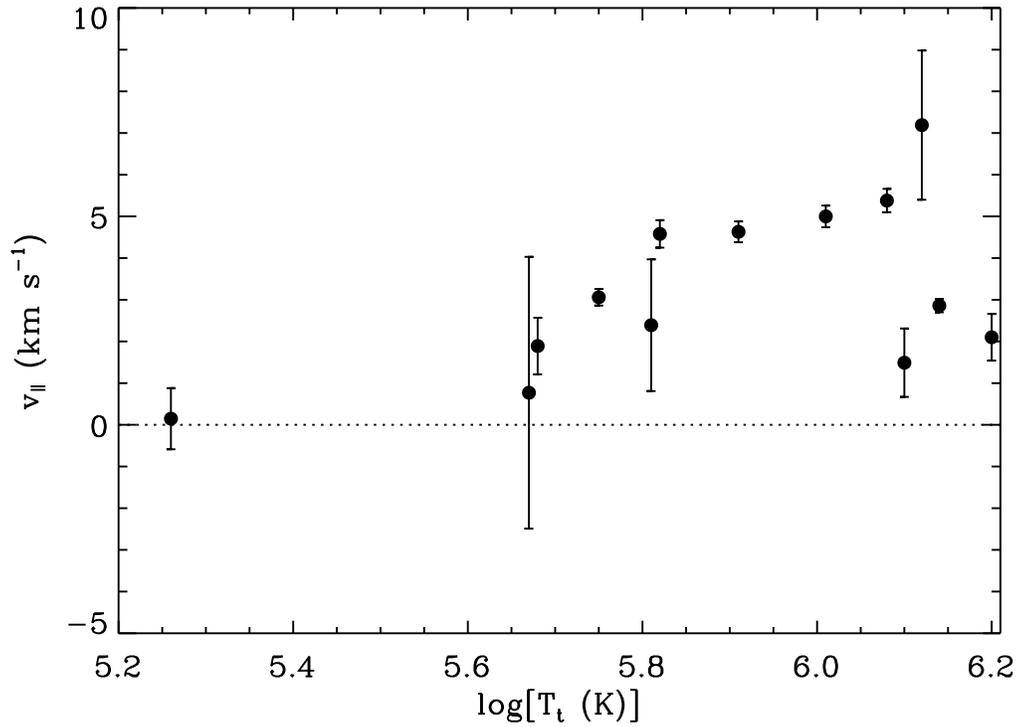}
\caption{\label{fig:outflow} Inferred flow velocity $v_{\parallel}$ along the magnetic field as a function of the DEM-averaged formation temperature $\log T_{t}$. The dotted horizontal line shows $v_{\parallel}=0$. There is a possible systematic offset in $v_{\parallel}$ of $\pm 1.3$~$\mathrm{km\,s^{-1}}$ due to the uncertainty in correcting for the tilt of the EIS slits relative to the CCD (see text for more detail). 
}
\end{figure}

	Figure~\ref{fig:outflow} plots the inferred $v_{\parallel}$ as a function of the DEM-averaged formation temperature $\log T_{t}$. For the analysis we performed the fit to equation~(\ref{eq:doppler}) for each line separately to determine $v_{\parallel}$ and then within each ion species took the weighted mean and error. The data point at $\log T_{t} = 5.26$ comes from the O~\textsc{iv} line, which is not used in the line width analysis. Since these lines come from the transition region it could be that a two-Gaussian fit would make a better fit, however the estimated systematic error from using a single Gaussian to fit these lines is $\sim 0.5$~$\mathrm{km\,s^{-1}}$ \citep{Peter:ApJ:1999}. 
	
	There is a systematic uncertainty in $v_{\parallel}$ due to the variation of the detector wavelength scale as a function of the Solar-Y position. This is caused by the tilt of the EIS slits relative to the CCD. We corrected for this variation using the parameters given by \citet{Kamio:SolPhys:2010}. However, due to the uncertainty in the fitting parameters, there is a possible $1\sigma$ systematic offset in the derived $v_{\parallel}$ of about $1.3$~$\mathrm{km\,s^{-1}}$ \citep{Young:EIS:2010}. 
	
	The results show that at coronal hole temperatures of $\log T_{\mathrm{e}} \approx 5.8$ -- $6.1$ there is an outflow velocity of $\approx 5$~$\mathrm{km\,s^{-1}}$. The flow velocity is small at both lower and higher temperatures. The small velocity at low temperatures implies that the outflow is smaller in the transition region. There may be two reasons for this. First, in the transition region there could be low lying cool closed loops which have no net outflow. Second, if temperature is proportional to height, then the increasing $v_{\parallel}$ with temperature may reflect acceleration within the transition region. The low velocity at high temperatures could be another indication that there are multiple structures within the large spatial bins and that the warm structures do not have strong outflows. 
	
	These results can be compared to those of \citet{Peter:ApJ:1999} and \citet{Tian:ApJ:2010}. \citet{Peter:ApJ:1999} measured Doppler velocities at disk center and found outflow velocities of $\approx 5$~$\mathrm{km\,s^{-1}}$ for $\log T_{\mathrm{e}} \gtrsim 6.0$, in reasonable agreement with the present measurements. However, they also found red-shifts at lower temperatures, in contrast to our results which show no flows at low temperatures. The reason for the difference could be that we are observing open field lines while the disk center observations of \citet{Peter:ApJ:1999} are dominated by closed field lines. Material on the open field lines is free to escape, while the trapped material on the closed field lines may return to the chromosphere as part of a mass cycle as described by \citet{McIntosh:ApJ:2012}. 
	
	\citet{Tian:ApJ:2010} measured Doppler shifts from a polar coronal hole and found that the outflow velocity increased steadily with formation temperature, reaching $\sim 25$~$\mathrm{km\,s^{-1}}$ at temperatures of $\log T_{\mathrm{e}}=6.2$. In our analysis we find much smaller velocities that are constant over a broad temperature range. One possible reason for the discrepancy is that \citet{Tian:ApJ:2010} focussed specifically on regions with a strong outflow. In the present analysis we performed no such selection. Since our method involves fitting over a range of $\alpha$ it necessarily finds an average outflow velocity for the coronal hole and could miss isolated regions of large outflow. The broad binning might also reduce the inferred velocity if such velocities were confined to small regions. Another possible reason is that the observation by \citet{Tian:ApJ:2010} looked near the boundary of the coronal hole, where there might be factors that increase the outflow, such as reconnection between the open and closed field lines. 
	
	Our measured velocities are also much smaller than the velocities of propagating coronal disturbances found in an equatorial coronal hole by \citet{McIntosh:SSR:2012}. They used sequences of Atmospheric Imaging Assembly (AIA) images to identify propagating disturbances moving with apparent velocities of $\sim 75$~$\mathrm{km\,s^{-1}}$. Such intermittent events would appear as line asymmetries in our data, but as discussed above (Section~\ref{subsec:fit}) we did not find any significant line asymmetries. If these jets occur relatively infrequently or cover small spatial scales, then they would make up only a small percentage of the total emission. It is possible that we do not resolve the asymmetries due to insufficient signal to noise or the large spatial integration in our data.
	
\section{Conclusion}\label{sec:con}

	We have described a new technique for combining magnetic field and spectroscopic data in order to derive plasma and wave properties in a coronal hole. We used a PFSS magnetic field model to determine the inclination angle between the line of sight and the magnetic field line in each spatial bin of our EIS spectrum. From these data we were able to infer the anisotropy of the ion temperatures, the non-thermal velocity, and the Doppler shift of plasma flowing along the field. The method provides both new results and also confirms other results that have been obtained using different methods.

	From the anisotropic line widths we inferred ion temperatures and non-thermal velocities induced by both transverse waves (Alfv\'en) and longitudinal waves (slow magnetosonic). The extracted $v_{\mathrm{nt},\perp}$ was consistent with estimates from off-limb observations for the Alfv\'en wave amplitude and $v_{\mathrm{nt},\parallel}$ was consistent with observations of quasi-periodic intensity oscillations seen in coronal holes. The inferred $T_{\mathrm{i},\perp}$ showed charge to mass ratio dependent heating, but with evidence suggesting greater collisional cooling than is found in off-limb measurements. The parallel ion temperature is expected to be equal to the proton temperature, for which there is limited data in the low corona. A comparison of our results for $T_{\mathrm{i},\parallel}$ with proton temperature measurements at slightly lower heights than this observation implies a large proton temperature gradient around 1.02~$R_{\sun}$. Our Doppler shift results indicate an average outflow velocity of about 5~$\mathrm{km\,s^{-1}}$ in the coronal hole over a broad temperature range. 
	
	One limitation for the current data set was the large spatial binning that was required in order to obtain good statistics for fitting the line profiles. Our analysis shows evidence that some of the material in these bins comes from warmer than average structures within the coronal hole. An observation with higher spatial resolution and better statistics may resolve some of these systematic issues. Similar techniques could also be applied to structures other than coronal holes. For example, nanoflare models predict anisotropic line broadening \citep[][]{Patsourakos:ApJ:2006} which might be observed in active regions or coronal loops. Such an analysis would likely require a more sophisticated magnetic field model than a PFSS.

\acknowledgements
We thank Enrico Landi for helpful suggestions and comments on the manuscript.
MH and DWS were supported in part by the NASA Solar Heliospheric Physics 
program grant NNX09AB25G and the NSF Division of Atmospheric and Geospace Sciences SHINE program
grant AGS-1060194. 
 	

\clearpage

\begin{center}
\begin{longtable}{llc c rcl}
\caption{Line List. \label{table:linelist}} \\
& Ion & & $\lambda$ (\AA)\tablenotemark{1} & \multicolumn{3}{c}{Transition\tablenotemark{1}} \\ \hline
\hline
\endfirsthead
& Ion & & $\lambda$ (\AA)\tablenotemark{1} & \multicolumn{3}{c}{Transition\tablenotemark{1}} \\ \hline
\hline
\endhead
\hline
\endlastfoot	
& & &256.317 & $1s\, ^{2}S_{1/2}$ & $-$ & $3p\, ^{2}P_{3/2}$ \\
&\raisebox{2.5ex}[0pt]{He \textsc{ii}} & \raisebox{2.5ex}[0pt]{$\Big\{$} & 256.318 & $1s\, ^{2}S_{1/2}$ & $-$ & $3p\, ^{2}P_{1/2}$ \\
&O \textsc{iv} & &279.631 & $2s^2\,2p\, ^{2}P_{1/2}$ & $-$ & $2s^2\,3s\, ^{2}S_{1/2}$ \\
&O \textsc{iv} & &279.933 & $2s^2\,2p\, ^{2}P_{3/2}$ & $-$ & $2s^2\,3s\, ^{2}S_{1/2}$ \\
&Mg \textsc{v} & &276.579 & $2s^2\, 2p^4\, ^{1}D_{2}$ & $-$ & $2s\, 2p^5\, ^{1}P_{1}$ \\
$\ast$&Mg \textsc{vi} & &268.991 &  $2s^2\, 2p^3\, ^{2}D_{3/2}$ & $-$ & $2s\, 2p^4\, ^{2}P_{1/2}$ \\
	& & &270.391 & $2s^2\, 2p^3\, ^{2}D_{5/2}$ & $-$ & $2s\, 2p^4\, ^{2}P_{3/2}$ \\
\raisebox{2.5ex}[0pt]{$\ast$}&\raisebox{2.5ex}[0pt]{Mg \textsc{vi}} &\raisebox{2.5ex}[0pt]{$\Big\{$} &270.400 & $2s^2\, 2p^3\, ^{2}D_{3/2}$ & $-$ & $2s\, 2p^4\, ^{2}P_{3/2}$ \\ 
$\ast$&Mg \textsc{vii} & &276.154 & $2s^2\, 2p^2\, ^{3}P_{0}$ & $-$ & $2s\, 2p^3\, ^{3}S_{1}$ \\
$\ast$&Si \textsc{vi} & &246.003& $2s^2\, 2p^5\, ^{2}P_{3/2}$ & $-$ & $2s\, 2p^{6}\, ^{2}S_{1/2}$ \\
$\ast$&Si \textsc{vi} & &249.125 & $2s^2\, 2p^5\, ^{2}P_{1/2}$ & $-$ & $2s\,2p^6\,^{2}S_{1/2}$ \\
$\ast$&Si \textsc{vii} & &272.648 & $ 2s^2\, 2p^4\, ^{3}P_2 $ &$-$ & $2s\, 2p^5\, ^{3}P_1 $ \\
$\ast$&Si \textsc{vii} & &275.361 & $ 2s^2\, 2p^4\, ^{3}P_2 $ &$-$ & $2s\, 2p^5\, ^{3}P_2 $ \\
$\ast$&Si \textsc{vii} & &275.676 & $ 2s^2\, 2p^4\, ^{3}P_1 $ &$-$ & $2s\, 2p^5\, ^{3}P_1 $ \\
$\ast$&Si \textsc{x} & &258.374 & $ 2s^2\, 2p\, ^2P_{3/2} $ &$-$ & $2s\, 2p^2\, ^{2}P_{3/2} $ \\
$\ast$&Si \textsc{x} & &261.057 & $2s^2\, 2p\, ^2P_{3/2} $ &$-$ & $2s\, 2p^2\, ^{2}P_{1/2} $ \\
$\ast$&Si \textsc{x} & &271.992 & $ 2s^2\, 2p\, ^2P_{1/2} $ &$-$ & $2s\, 2p^2\, ^{2}S_{1/2} $ \\
$\ast$&Si \textsc{x} & &277.264 & $ 2s^2\, 2p\, ^{2}P_{3/2} $ &$-$ & $ 2s\, 2p^2\, ^{2}S_{1/2} $ \\
$\ast$&S \textsc{x}  & &264.231 & $ 2s^2\, 2p^3\, ^{4}S_{3/2}$ &$-$ & $2s\, 2p^4\, ^{4}P_{5/2} $ \\
$\ast$&Fe \textsc{viii} & &185.213 & $ 3p^6\, 3d\, ^{2}D_{5/2} $ &$-$ & $3p^5\,3d^2\, (^{3}F)\, ^{2}F_{7/2}$ \\
$\ast$&Fe \textsc{viii} & & 186.599 & $3p^6 \,3d\, ^{2}D_{3/2} $ &$-$ & $3p^5\, 3d^2\, (^{3}F)\, ^{2}F_{5/2}$ \\
$\ast$&Fe \textsc{viii} & & 194.661 & $ 3p^6 \, 3d\, ^{2}D_{5/2} $ &$-$ & $3p^6\, 4p\, ^{2}P_{3/2}$ \\
$\ast$&Fe \textsc{ix} & &188.497 & $ 3s^2\, 3p^5\, 3d\, ^{3}F_4 $ &$-$ & $3s^2\, 3p^4\, (^{3}P)\, 3d^2\, ^{3}G_5 $\\
$\ast$&Fe \textsc{ix} & &189.941 & $ 3s^2\, 3p^5\, 3d\, ^{3}F_3 $ &$-$ & $3s^2\, 3p^4\, (^{3}P)\, 3d^2\, ^{3}G_4 $\\
$\ast$&Fe \textsc{ix} & &197.862 & $ 3s^2\, 3p^5\, 3d\, ^{1}P_1 $ &$-$ & $3s^2\, 3p^5\, 4p\, ^{1}S_0 $ \\
&Fe \textsc{x} & &174.531 & $3s^2\, 3p^5\, ^{2}P_{1/2} $ &$-$ & $3s^2\, 3p^4\, (^{3}P)\, 3d\, ^{2}D_{5/2} $\\
$\ast$&Fe \textsc{x} & &184.537 & $ 3s^2\, 3p^5\, ^{2}P_{3/2} $ &$-$ & $3s^2\, 3p^4\, 	(^1D)\, 3d\, ^{2}S_{1/2} $ \\
$\ast$&Fe \textsc{x} & &190.037 & $ 3s^2\, 3p^5\, ^{2}P_{1/2} $ &$-$ & $3s^2\, 3p^4\, (^{1}D)\, 3d\, ^{2}S_{1/2} $ \\
&Fe \textsc{x} & &193.715 & $ 3s^2\, 3p^5\, ^{2}P_{3/2} $ &$-$ & $3s^2\, 3p^4\, (^{1}S)\, 3d\, ^{2}D_{5/2} $ \\
														&	&	& 257.259 & $ 3s^2\, 3p^5\, ^{2}P_{3/2} $ &$-$ & $3s^2\, 3p^4\, (^{3}P)\, 3d\, ^{4}D_{5/2} $ \\*
\raisebox{2.5ex}[0pt]{$\ast$}&\raisebox{2.5ex}[0pt]{Fe \textsc{x}}&\raisebox{2.5ex}[0pt]{$\Big\{$} & 257.263 & $3s^2\, 3p^5\, ^{2}P_{3/2}$ & $-$ & $3s^2\, 3p^4\, (^{3}P)\, 3d\, ^{4}D_{7/2}$\\
&Fe \textsc{xi} & &180.401 & $ 3s^2\, 3p^4\, ^{3}P_2 $ &$-$ & $3s^2\, 3p^3\, (^{4}S)\, 3d\, ^{3}D_3 $\\
&Fe \textsc{xi} & &182.167 & $ 3s^2\, 3p^4\, ^{3}P_1 $ &$-$ & $3s^2\, 3p^3\, (^{4}S)\, 3d\, ^{3}D_2 $ \\
$\ast$&Fe \textsc{xi} & &188.217 & $ 3s^2\, 3p^4\, ^{3}P_2 $ &$-$ & $3s^2\, 3p^3\, ( ^{2}D)\, 3d\, ^{3}P_2 $ \\
$\ast$&Fe \textsc{xi} & &188.299 & $ 3s^2\, 3p^4\, ^{3}P_2 $ &$-$ & $3s^2\, 3p^3\, (^{2}D)\, 3d\, ^{1}P_1 $ \\
&Fe \textsc{xi} & &189.711 & $3s^2\, 3p^4\, ^{3}P_{0}$ &$-$ & $3s^2\, 3p^3\, (^{2}D)\, 3d\, ^{3}P_{1}$ \\
$\ast$&Fe \textsc{xii} & &192.394 & $3s^2\, 3p^3\, ^{4}S_{3/2}$ &$-$ & $3s^2\, 3p^2\, (^{3}P)\, 3d\, ^{4}P_{1/2}$ \\
&Fe \textsc{xii} & &193.509 & $3s^2\, 3p^3\, ^{4}S_{3/2}$ &$-$ & $3s^2\, 3p^2\, (^{3}P)\, 3d\, ^{4}P_{3/2}$ \\
 & 	& &195.119 & $ 3s^2\, 3p^3\, ^{4}S_{3/2} $ &$-$ & $3s^2\, 3p^2\, (^{3}P)\, 3d\, ^{4}P_{5/2} $ \\
\raisebox{2.5ex}[0pt]{$\ast$}&\raisebox{2.5ex}[0pt]{Fe \textsc{xii}} &\raisebox{2.5ex}[0pt]{$\Big\{$} &195.179 & $ 3s^2\, 3p^3\, ^{2}D_{3/2} $  &$-$ & $3s^2\, 3p^2\, (^{1}D)\, 3d\, ^{2}D_{3/2} $ \\
$\ast$&Fe \textsc{xiii} & &202.044 & $ 3s^2\, 3p^2\, ^{3}P_0 $ &$-$ & $3s^2\, 3p\, 3d\, ^{3}P_1 $ \\
		&	&  &203.772 & $ 3s\, 3p^3\, ^{3}D_{1}$ &$-$ & $ 3s, 3p^2\, 3d\, ^{3}F_{2}$ \\*
    &  & &203.796 & $ 3s^2\, 3p^2\, ^{3}P_{2}$ &$-$ & $ 3s^2\, 3p\, 3d\, ^{3}D_{2}$ \\*
		&  & &203.827 & $ 3s^2\, 3p^2\, ^{3}P_{2}$ &$-$ & $ 3s^2\, 3p\, 3d\, ^{3}D_{3}$ \\*
&\raisebox{7.0ex}[0pt]{Fe \textsc{xiii}}	&\raisebox{7.0ex}[0pt]{$\left\{\rule{0cm}{1.2cm}\right.$} &203.835 & $ 3s\, 3p^3\, ^{3}D_{2}$ &$-$ & $ 3s, 3p^2\, 3d\, ^{3}F_{2}$ \\		
&Fe \textsc{xiv} & &270.521 & $3s^2\, 3p\, ^{2}P_{3/2}$ & $-$ & $3s\, 3p^{2}\, ^{2}P_{1/2}$ \\
&Fe \textsc{xv} & &284.163 & $3s^2\, ^{1}S_{0}$ & $-$ & $3s\, 3p\, ^{1}P_{1}$ 
\footnotetext[1]{Wavelengths and transitions taken from CHIANTI \citep{Dere:AA:1997, Landi:ApJ:2012}.}
\footnotetext[0]{Brackets indicate blends from the same ion. Asterisks mark the lines used in the analysis of the line widths. These and the remaining lines were all used for the DEM analysis.}
\end{longtable}
\end{center}

\begin{table}
\centering
\caption{Line Widths and Effective Formation Temperatures \label{table:components}}
\begin{tabular}{lllll}
& Ion 
& $v_{\mathrm{eff},\parallel}$~$(\mathrm{km\,s^{-1}})$ 
&	$v_{\mathrm{eff},\perp}$~$(\mathrm{km\,s^{-1}})$ 
& $\log [T_{t} (\mathrm{K})]$ \\
\hline
&Mg~\textsc{vi} 	&	$36.1 \pm 1.7$ 		& $33.6 \pm 1.0$ 	& 	$5.68$ \\
&Mg~\textsc{vii} 	& $43.2 \pm 5.4$ 		& $31.4 \pm 1.8$ 	& 	$5.81$ \\
&Si~\textsc{vi} 	&	$33.8 \pm 8.0$ 		& $40.0 \pm 3.7$ 	& 	$5.67$ \\
&Si~\textsc{vii} 	& $35.1 \pm 0.8$ 		& $36.2 \pm 0.3$ 	& 	$5.82$ \\
&Si~\textsc{x}		& $30.7 \pm 2.5$ 		& $36.6 \pm 1.2$	& 	$6.10$ \\
&S~\textsc{x} 		& $42.6 \pm 6.4$ 		& $35.0 \pm 2.8$ 	& 	$6.12$ \\
&Fe~\textsc{viii}	& $26.5 \pm 0.7$		& $32.2 \pm 0.3$	& 	$5.75$ \\
&Fe~\textsc{ix}		& $25.5 \pm 1.1$		& $34.1 \pm 0.4$	& 	$5.91$ \\
&Fe~\textsc{x}		& $27.2 \pm 1.0$ 		& $32.4 \pm 0.4$	& 	$6.01$ \\
&Fe~\textsc{xi}		& $24.2 \pm 0.9$		& $29.1 \pm 0.4$	& 	$6.08$ \\
&Fe~\textsc{xii}	& $27.1 \pm 0.7$		& $30.7 \pm 0.3$	& 	$6.14$ \\
&Fe~\textsc{xiii}	& $23.2 \pm 2.3$		& $27.5 \pm 1.0$	& 	$6.20$ \\
\hline
\end{tabular}
\end{table}

\bibliography{EquatorialCH}

\end{document}